\documentclass[%
 reprint,
superscriptaddress,
 amsmath,amssymb,
 aps,
pra,
]{revtex4-2}

\usepackage{graphicx}
\usepackage{dcolumn}
\usepackage{bm}

\begin{document}

\preprint{APS/123-QED}

\title{Long-distance free-space quantum key distribution with continuous variables}

\author{Tianxiang Zhan}
\affiliation{%
 State Key Laboratory of Photonics and Communications, Institute for Quantum Sensing and Information Processing, Shanghai Jiao Tong University, Shanghai 200240, China
}%

\author{Huasheng Li}
\affiliation{%
 Shanghai XunTai Quantech Co., Ltd, Shanghai, 200241, China
}%

\author{Peng Huang}
 \email{Corresponding author: huang.peng@sjtu.edu.cn}
\affiliation{%
 State Key Laboratory of Photonics and Communications, Institute for Quantum Sensing and Information Processing, Shanghai Jiao Tong University, Shanghai 200240, China
}%
\affiliation{
 Shanghai Research Center for Quantum Sciences, Shanghai 201315, China
}%
\affiliation{
 Hefei National Laboratory, Hefei 230088, China
}

\author{Haoze Chen}
\affiliation{%
 CAS Quantum Network Co., Ltd., Shanghai, 200345, China
}%

\author{Jiaqi Han}
\affiliation{%
 Shanghai XunTai Quantech Co., Ltd, Shanghai, 200241, China
}%

\author{Zijing Wu}
\affiliation{%
 Shanghai XunTai Quantech Co., Ltd, Shanghai, 200241, China
}%

\author{Hao Fang}
\affiliation{%
 Shanghai XunTai Quantech Co., Ltd, Shanghai, 200241, China
}%

\author{Hanwen Yin}
\affiliation{%
 State Key Laboratory of Photonics and Communications, Institute for Quantum Sensing and Information Processing, Shanghai Jiao Tong University, Shanghai 200240, China
}%

\author{Zehao Zhou}
\affiliation{%
 State Key Laboratory of Photonics and Communications, Institute for Quantum Sensing and Information Processing, Shanghai Jiao Tong University, Shanghai 200240, China
}%

\author{Huiting Fu}
\affiliation{%
 College of Information Science and Technology, Donghua University, Shanghai 201620, People’s Republic of China
}%

\author{Feiyu Ji}
\affiliation{%
 State Key Laboratory of Photonics and Communications, Institute for Quantum Sensing and Information Processing, Shanghai Jiao Tong University, Shanghai 200240, China
}%

\author{Piao Tan}
\affiliation{%
 State Key Laboratory of Photonics and Communications, Institute for Quantum Sensing and Information Processing, Shanghai Jiao Tong University, Shanghai 200240, China
}%

\author{Yingming Zhou}
\affiliation{%
 Shanghai XunTai Quantech Co., Ltd, Shanghai, 200241, China
}%

\author{Xueqin Jiang}
\affiliation{%
 College of Information Science and Technology, Donghua University, Shanghai 201620, People’s Republic of China
}%

\author{Tao Wang}
\affiliation{%
 State Key Laboratory of Photonics and Communications, Institute for Quantum Sensing and Information Processing, Shanghai Jiao Tong University, Shanghai 200240, China
}%
\affiliation{
 Shanghai Research Center for Quantum Sciences, Shanghai 201315, China
}%
\affiliation{
 Hefei National Laboratory, Hefei 230088, China
}

\author{Jincai Wu}
\affiliation{%
 Key Laboratory of Space Active Optoelectronic Technology, Shanghai Institute of Technical Physics, Chinese Academy of Sciences, Shanghai, 200083, China
}%

\author{Cheng Ye}
\affiliation{%
 CAS Quantum Network Co., Ltd., Shanghai, 200345, China
}%

\author{Yajun Miao}
\affiliation{%
 CAS Quantum Network Co., Ltd., Shanghai, 200345, China
}%

\author{Wei Qi}
\affiliation{%
 CAS Quantum Network Co., Ltd., Shanghai, 200345, China
}%

\author{Guihua Zeng}
 \email{Corresponding author: ghzeng@sjtu.edu.cn}
\affiliation{%
 State Key Laboratory of Photonics and Communications, Institute for Quantum Sensing and Information Processing, Shanghai Jiao Tong University, Shanghai 200240, China
}%
\affiliation{%
 Shanghai XunTai Quantech Co., Ltd, Shanghai, 200241, China
}%
\affiliation{
 Shanghai Research Center for Quantum Sciences, Shanghai 201315, China
}%
\affiliation{
 Hefei National Laboratory, Hefei 230088, China
}

\date{\today}

\begin{abstract}
Continuous-variable quantum key distribution (CVQKD) enables remote users to share high-rate and unconditionally secure secret keys while maintaining compatibility with classical optical communication networks and effective resistance against background noise. However, CVQKD experiments have only been demonstrated indoors or over short outdoor distances. Here, by developing channel-fluctuation-independent high-precision manipulation of continuous-variable quantum states, high-accuracy quantum signal acquisition and processing, and high-efficiency free-space acquisition, tracking, and pointing technology, we overcome the excess noise due to atmospheric effects especially in daylight without extra wavelength conversion and spectral filtering, and demonstrate for the first time long-distance free-space quantum key distribution over 7-km inland and 9.6-km maritime atmospheric channels with Gaussian-modulated coherent states. This achieved distribution distance of secure quantum secret keys is well beyond the atmosphere's effective thickness, offering a promising alternative for realizing satellite-based quantum cryptography communication in daylight. Moreover, given that the CVQKD system is naturally compatible with existing ground fiber telecommunication networks, it marks an essential step for realizing integrated air-ground quantum access networks with cross-domain applications. 
\end{abstract}

\maketitle



\textit{Introduction}.\textemdash Continuous-variable quantum key distribution (CVQKD) \cite{TC99, GG02, GG03, CS12,yc24} enables exchange of keys between remote communicating parties with information-theoretic security based on the principle of quantum physics. The secret key information is usually encoded in the quadratures \cite{GG02,ptwo08, AL09,tm21} of the quantized electromagnetic field of coherent states with standard telecommunication components, 
and detected by coherent receiver, such as homodyne or heterodyne detection. So, CVQKD inherits the advantages of cost-effective and high-speed solutions of coherent optical communication to realize high secure-key rates. Moreover, the ideal implementation of CVQKD can approximately reach the the PLOB bound \cite{PLOB17}, which gives the ultimate limit of secret key capacity of repeaterless quantum communication. Until now, Gaussian-modulated coherent-state (GMCS) CVQKD protocols \cite{GG02,ptwo08, AL09,tm21} have been progressively proven to be secure against collective and coherent attacks without \cite{COL06, COR09} and with \cite{al10, FF12, FIN13, FIN17, COM15, COM18} considerations of finite-size effects. 

So far, the feasibilities of free-space and satellite-based CVQKD have been theoretically and experimentally investigated \cite{usenko2012, vas12, sem12, sy18, papa18, wang2019, LR2019, hos2019, DL2021}, and 
secure key distributions have been demonstrated through the real outdoor free-space atmospheric channels with polarization \cite{elser09, shi19, Zheng2025} and quadrature \cite{ivan16} encodings. However, these reported outdoor free-space CVQKD experiments were demonstrated over a distance of up to 1.6 km \cite{ivan16}. Especially as the transmission distance becomes longer, the transmission attenuation and its fluctuation, will severely increase, and the induced changes of intensity, phase, and polarization of quantum signals will become more intense. Therefore, it will be difficult to precisely acquire and rebuild the encoded quadrature component, experienced transmission attenuation, and related shot noise for each quantum state, which will inevitably induce large excess noise.

\begin{figure*}[!htbp]
	\centering
	\includegraphics[width=0.95\textwidth]{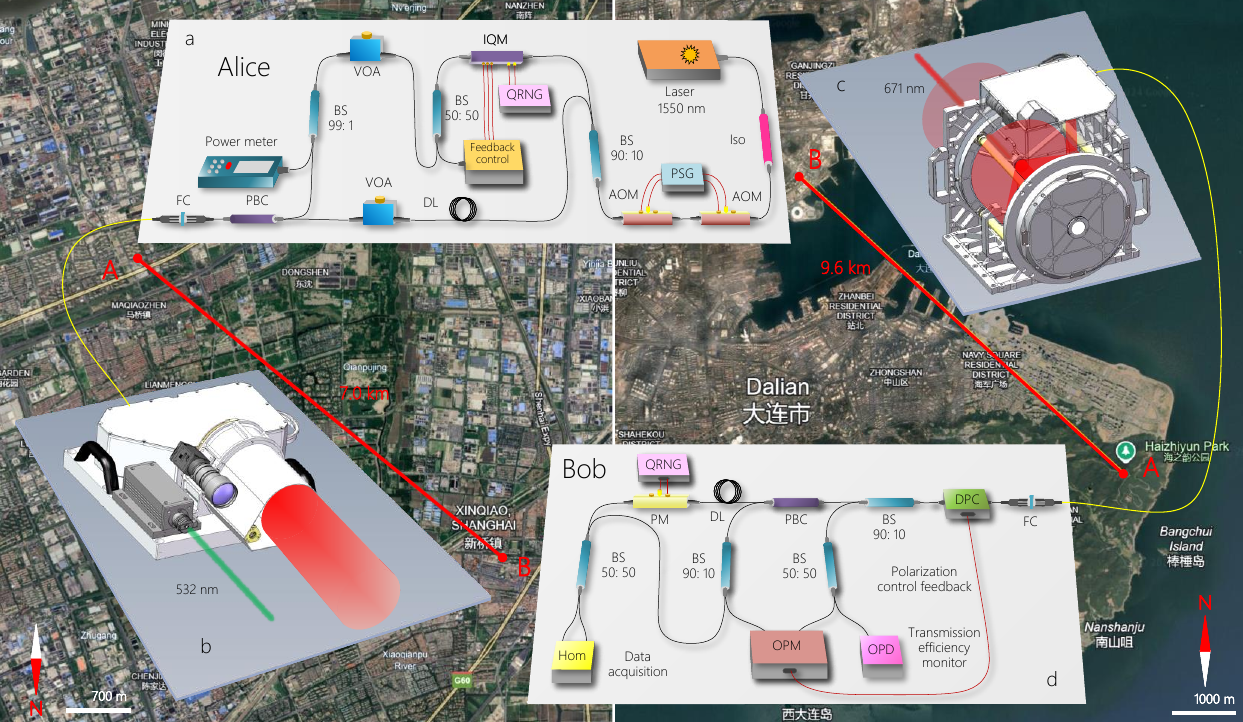}
	\caption{(Color online). 
    All-day free-space CVQKD experiments over 7-km inland atmospheric channel (left) and 9.6-km maritime atmospheric channel (right). a, The free-space CVQKD transmitter; b, the transmitting telescope; c, the receiving telescope; d, The free-space CVQKD receiver. BS, beam splitter; PSG, pulse signal generator; IQM, in-phase and quadrature modulator; QRNG, quantum random number generator; VOA, variable optical attenuators; DL, delay line; PBC, polarization beam combiner; SMF, single-mode fiber; OPM, optical power monitor; OPD, optical power detector; ADC, analog-to-digital converter; PM, phase modulator; HOM, homodyne detector.
    Map data: Google, Airbus Maxar Technologies, 2023/6/10.}\label{exp1}
\end{figure*}

Here, we 
report a GMCS CVQKD system capable of generating all-day secret keys against coherent attacks asymptotically over 7-km inland and 9.6-km maritime atmospheric channels. The maximum subchannel secure key rate, including the daytime and nighttime, can exceed 700 bps for field-test scenarios. In particular, the secret keys are generated by controlling the excess noise and transmission loss in three ways. First, we develop high-precision manipulation of continuous-variable quantum states to correct the drifted polarization and phase independent on the severe fluctuation of transmission efficiency.
Second, we propose synchronization-trigger sampling and data-aided signal processing methods for high-accuracy acquisitions of quantum key information and real-time shot noise, so as to realize low-noise grouped parameter estimations regardless of intense decrease and variation of transmittance. Finally, we adopt a high-efficiency free-space acquisition, tracking, and pointing (ATP) technology to control the transmission attenuation of quantum signals and improve the signal-to-noise ratio (SNR) in the secret key generation area. 
Altogether, our free-space CVQKD system enables transmission distance enhancement of about half an order of magnitude greater than the previous record \cite{ivan16}. The system's effectiveness and robustness are verified by all-day operation, which presents an essential step for realizing integrated air-ground quantum networks with applications of all-day secure communications. 

\textit{Scheme description}.\textemdash Our free-space CVQKD system is based on GMCS protocol with homodyne detection \cite{GG02}. 
Alice generates $2N$ real random variables $X_{A}^{1},\dots,X_{A}^{N}, P_{A}^{1},\dots,P_{A}^{N}$ which follows Gaussian distribution $X_A^1, \dots, X_A^N, P_A^1, \dots, P_A^N \stackrel{\text{i.i.d.}}{\sim} \mathcal{N}(0, V_A)$,
and prepares correspondingly $N$ coherent states $|\alpha_{1}\rangle,\dots,|\alpha_{N}\rangle$ with $\alpha_{k}=X_{A}^{k}+iP_{A}^{k}\in\mathbb{C}$. Each coherent state is sent through the quantum channel, where Bob randomly measures the X or P quadrature via homodyne detection. 
Alice and Bob then perform quadrature basis alignment and sift $N$ pairs of correlated variables. For the case of reverse reconciliation, the sifted measurement results form the raw key set $Y$, which will be used to further distill the secret keys with reconciliation and privacy amplification operations. The asymptotic secret key rate ($N\rightarrow\infty$) with reverse reconciliation can be obtained with the Devetak-Winter bound \cite{dev05}
\begin{equation}\label{e1}
R=f(1-\alpha)(1-FER)(\beta I_{AB}-\chi_{BE}),
\end{equation}
where $f$ is the symbol rate for quantum signal, $I_{AB}$ is the classical mutual information between Alice and Bob, $\chi_{BE}$ is the Holevo bound on the information leaked to Eve, $\alpha$ is the overhead for frame synchronization and parameter estimation, $\beta$ and $\text{FER}$ are the efficiency and frame error rate of the reconciliation, respectively. 

Different from the fiber channel with supposed constant attenuation, the transmission efficiency $T^{k}$ of free-space link fluctuates due to atmospheric effects with the typical order of 1-10 kHz frequency of fading process \cite{usenko2012, DL2021}. We can find the transmission efficiency fluctuates severely over $N$ uses, making the receiving state a mixture of the individual fixed-attenuation states and resulting in a non-negligible extra excess noise in parameter estimation \cite{usenko2012}. By monitoring the current transmission loss of the quantum channel, Alice and Bob can sort the shared raw key data into different groups $g_{i} (i=1,\dots, L)$ according to divided channel transmission efficiency interval $[T_{i}-\Delta T/2, T_{i}+\Delta T/2]$, in which the corresponding transmission efficiencies $T_{i}^{k}$ are approximately equal and just with slight fluctuations, to significantly relive the extra fading excess noise \cite{LR2019}. The total secret key rate of free-space CVQKD can be expressed as
\begin{equation}\label{e2}
R_{tot}=\sum_{i=1}^{L}P(\{T_{i}^{k}\}\subseteq\{T^{k}\})\alpha_{G}(g_{i})R(g_{i}),
\end{equation}
where $P(\{T_{i}^{k}\}\subseteq\{T^{k}\})$ represents the probability of group $g_{i}$ with $\{T_{i}^{k}\}$ in the whole raw key data set, $\alpha_{G}(g_{i})$ represents the secure-key-generation proportion of raw key data within group $g_{i}$, and $R(g_{i})$ denotes the secret key rate of group $g_{i}$, which can be calculated with Eq.~(\ref{e1}) (see Supplementary Material for more details).

\textit{Implementation setup}.\textemdash
The system depicted in Fig.~\ref{exp1} is built to implement the scheme. The CVQKD experiments are conducted via an inland atmospheric channel in Shanghai, China, from April 29 to July 8, 2024 and a maritime atmospheric channel in Dalian, China, from October 21 to November 16, 2024. For the inland atmospheric environment, the sending terminal (Alice) is located at Minhang District (N31$^{\circ}$01$'$23$''$, E121$^{\circ}$$22'08''$), Shanghai City. 
Alice sends the 1550 nm optical signals through a 7 km free-space link to the receiving terminal (Bob), which is located at Songjiang District (N31$^{\circ}$$03'46''$, E121$^{\circ}$$18'46''$), Shanghai City. For the maritime atmospheric environment, the sending terminal and the receiving terminal are located in the Zhongshan District (N38$^{\circ}$54$'$02$''$, E121$^{\circ}$$42'29''$) and Ganjingzi District (N38$^{\circ}$57$'$32$''$, E121$^{\circ}$$37'31''$) of Dalian City, respectively, with a direct line-of-sight distance of 9.6 km, partly passing over the ocean surface of Dalian Bay.

Two cascaded acoustic optical modulators (AOM) are applied to cut the 1550-nm continuous-wave (CW) light into a 5-MHz high-stability and high-extinction-ratio coherent pulse train with 20\% duty cycle. 
These pulses are split into weak signals and LO paths. The key information is encoded in the X and P quadratures of the weak coherent pulses with an IQ modulator and further attenuated to proper modulation variance (Fig.~\ref{exp1}a). By employing the polarization-division multiplexing and time-division multiplexing method, the encoded quantum signals together with the LO are collimated out into free space by the sending telescope (mounted on a three-dimensional platform, Fig.~\ref{exp1}b). At the receiving terminal, the signal light is collected by a receiving telescope (Fig.~\ref{exp1}c) and goes through a 1-m long SMF to the receiving terminal (Fig.~\ref{exp1}d). A dynamic polarization controller (DPC) is used to calibrate the polarization states of the received pules to demultiplex the LO and signal pulses precisely. Then Bob randomly measures one of the quadratures with a home-made high-gain homodyne detector and outputs the results with a 1 GS/s data acquisition module.

For a practical atmospheric channel, 
controlling the untrusted excess noise and transmission attenuation arising from atmospheric effects is the core problem for secure key generation.
The untrusted excess noise that arises from different sources \cite{Laudenbach2018}, including the polarization leakage (PL) from transmitted LO, phase excess noise, ADC quantization noise, relative intensity noise (RIN) and so on, can be decomposed into independent individual contributions
\begin{equation}
\varepsilon=\varepsilon_{PL}+\varepsilon_{phase}+\varepsilon_{ADC}+\varepsilon_{RIN}+O,
\end{equation}
where the first four items are the main sources of excess noise in free-space CVQKD with transmitted LO, and $O$ is the other noise sources involving modulation, channel fading, and common mode rejection ratio (CMRR), etc. \cite{fc2018} (see Supplementary Material for more details).

\begin{figure}[!htbp]
	\centering
	\includegraphics[width=0.45\textwidth]{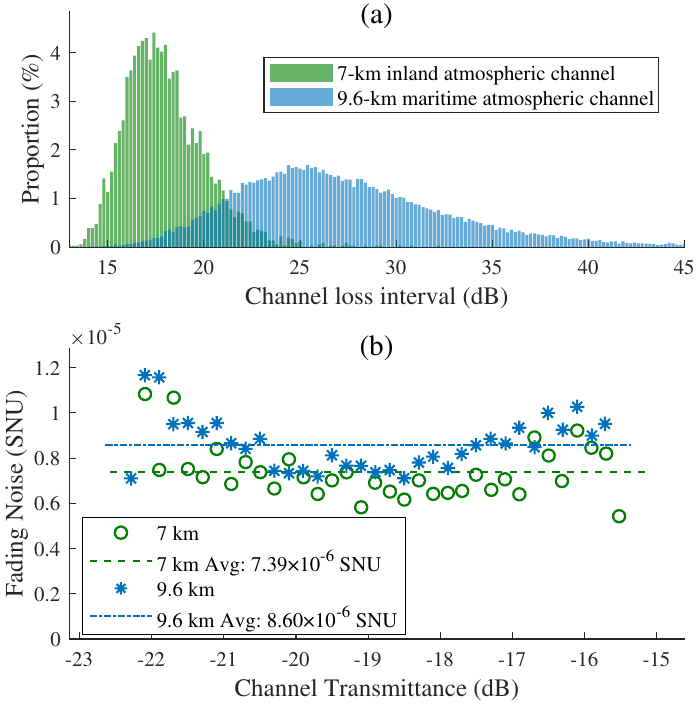}
	\caption{(Color online). Statistical Characteristics for free-space CVQKD system over 7-km and 9.6-km atmospheric channel. (a) The proportion in different 0.2-dB channel loss intervals. (b) The measured fading excess noises in different 0.2-dB channel loss intervals and the average values.
 }\label{ChanSta}                                                
\end{figure}

We develop channel-fluctuation-independent manipulation methods for precise restorations of the polarization and phase states of the encoded continuous-variable quantum signals to suppress the untrusted excess noise components $\varepsilon_{PL}$ and $\varepsilon_{phase}$. 
For the dynamic depolarization calibration, 
we innovatively split 9\% quantum signal and 5\% LO signals to calculate their ratio to provide feedback independent of channel loss fluctuation. The polarization extinction ratio can be regularly maintained between 30-40 dB, independent of the fluctuation of transmission efficiency
(see Supplementary Material for more details). 
For phase drift compensation, 
two adjacent high-intensity quadrature phase shift keying (QPSK) pilot pulses 
are employed to calibrate one quantum signal phase, enabling precise phase compensation for each signal and reducing phase drift to within the secure key generation area
(see Supplementary Material for more details).
Moreover, an ATP system with a high-efficiency free-space SMF coupling technique is developed, where a coarse-fine two-stage tracking system is designed to simultaneously implement higher feedback bandwidth and longer tracking stability (see Supplementary Material for more details). The whole transmission efficiencies can be controlled to less than 25 dB shown in Fig.~\ref{ChanSta}a. 

To accurately acquire the quantum key information after detection and perform parameter estimation to bound the secret key rate, the transmission efficiency is monitored by employing a 12-bit ADC to sample the relatively high-intensity output pulsed 5\% LO of OPD with 1 GS/s. One can judge and acquire the peak value in each cycle and obtain the experienced transmission efficiency for each pulse (including the quantum and pilot signals). 
Therefore, by calibrating the delay between the signal path and the split monitoring path in advance, the peak data of quantum signals can be precisely extracted independent on channel fluctuation. Meanwhile, the shot noise can also be obtained in real-time by randomly inserting vacuum states as quantum signals or using the precalibrated relationship between the monitored output value 
and shot noise. To reduce the effect of fading excess noise $\varepsilon_{f}$ \cite{usenko2012}, the quantum key data are uniformly divided into several groups after phase compensation according to the monitored probability distribution of the transmission efficiency (PDTE). Here, the PDTE interval width $\Delta T$ is set as 0.2 dB equally to compromise the fading excess noise to negligible values shown in Fig.~\ref{ChanSta}b.

\begin{table*}
	\centering
	\caption{\textbf{Experimental parameters summary of all-day free-space CVQKD over 7-km inland atmospheric channel and 9.6-km maritime atmospheric channel.} $\lambda$, quantum signal wavelength; $D_{T}$, emitting telescope diameter; $D_{R}$, receiving telescope diameter; $\theta_{p}$, pointing error; $\theta_{d}$, divergence angle; $\eta$, detection efficiency; $N.A.$, the numerical aperture; $d_{cor}$, fiber core diameter; $f$, symbol rate for the quantum signal; $L$, the average channel loss in this data group; $P$, the probability for the channel loss interval subset of this experiment; $V_{A}$, modulation variance; $\nu_{el}$, electronic noise; SNR$_{s}$, the signal-to-noise ratio of quantum signal; $\varepsilon$, excess noise; $\beta$, reconciliation efficiency; FER, frame error rate; $R$, secret key rate of different groups.
    The distinction between day and night data is based on the local sunset time. If the data acquisition period spans sunset time, the classification into ``day'' or ``night'' is determined by the predominant source of data used to establish the secure key successfully.}
		\begin{tabular}{c| c c c c c c c c c c}
			\hline
			\hline
			Fixed parameters & $\lambda$ & $D_{T}$ & $D_{R}$ & $\theta_{p}$ & $\theta_{d}$ & $\eta$ & N.A. &  $\alpha$ &  $d_{cor}$ &  $f$ \\
			Designed values  &1550 nm	&80 mm	&250 mm	&3 $\mu$rad& 32.7 $\mu$rad &0.375 & 0.125	& 0.5	&10 $\mu$m  & 2.5MHz\\
			\hline
			\hline
			Dynamical parameters & $L$  & $P$ & $V_{A}$ & $\nu_{el}$ & SNR$_{s}$ & $\varepsilon$ & $\alpha_{G}$ & $\beta$ &  FER &  $R$  \\
			\hline
			 & \multicolumn{10}{c}{The optimal group for each experiment over 7-km inland atmospheric channel} \\
			1st experiment (Night)&19.9018 dB&0.31\%&6.7433	&0.3748	&0.0188&0.0240 &12.50\% &95.0\%	&82\%	&130.2568 bps \\
			2nd experiment (Day) &22.1069 dB &2.03\%	&4.2718	&0.6314	&0.0060	&0.0487	&100.00\% &96.5\%	&88\%	&28.1797 bps \\
			3rd experiment (Night) &20.8929 dB&1.87\%&8.0596	&0.4587	&0.0169&0.0250	&16.67\% &96.5\%	&80\%	&132.1770 bps \\
			4th experiment (Night) &20.7102 dB&1.30\%&6.6863	&0.4526	&0.0147&0.0109	&12.50\% &96.5\%	&90\%	&98.4031 bps \\
			5th experiment (Night) &19.7104 dB &1.73\% &7.6867	&0.3488	&0.0228	&0.0411	&20.00\% &96.5\%	&79\%	&134.0639 bps\\
			6th experiment (Night)&16.5020 dB&0.05\%&8.9241	&0.1494	&0.0651 &0.0258	&33.33\% &96.5\%	&86\%	&334.5609 bps \\
			\hline
			 & \multicolumn{10}{c}{The optimal group for each experiment over 9.6-km maritime atmospheric channel} \\
			1st experiment (Day) &19.8899 dB &0.64\% &9.0989	&0.3186	&0.0265	&0.0035	&5.88\% &96.5\%	&48\%	&721.1802 bps \\ 
			2nd experiment (Day) &19.4962 dB &0.16\% &8.7341	&0.3017	&0.0283	&0.0133	&12.50\% &96.5\%	&43\%	&740.6015 bps\\
			3rd experiment (Day) &17.5291 dB&0.03\%&9.1478	&0.1972	&0.0506&0.0071	&9.09\% &96.0\%	&86\%	&319.8391 bps \\
			\hline
			\hline
		\end{tabular}
	\label{tab}
\end{table*}

\textit{Field-test results}.\textemdash
We calibrated the excess noise, modulation variances, quantum signal SNRs, channel loss, and the probabilities of corresponding channel loss intervals for different grouped data blocks.
The partial high-key-rate results for the grouped data blocks from each experiment over 7 km and 9.6 km are depicted in Table~\ref{tab} (see Supplementary Material for more experimental results). For the inland experiment, 
the channel loss at night is generally smaller than during the day, leading to the majority of experiments being conducted at night. However, the secure keys are also successfully generated during the daytime. 
In contrast to the inland atmospheric environment, the channel attenuation in the marine environment is generally lower at day than during the night. This difference is likely attributable to the unique climatic conditions of the maritime environment during that season, where temperatures decrease and humidity increases after sunset, leading to the emergence of turbulence and the formation of fog over the sea. Therefore, most experiments under the marine environment are conducted during the day, and a smaller proportion of data is collected at night. Nevertheless, secure keys are also successfully generated in the maritime atmospheric environment. 
In our experiments, the SNR ranges from 0.0060 to 0.0651. The multidimensional reconciliation \cite{la08,jp11} with the rate-adaptive low-density parity check (LDPC) codes is adopted to achieve low-SNR decoding at different SNRs to meet the decoding requirements with reconciliation efficiency 95\%-96.5\% and frame error rate (FER) 43\%-90\%. The achieved average secret key rates range from 0.0560 to 2.3545 bps over 7 km and from 0.0379 to 0.4768 bps over 9.6 km, as obtained using Eq.~(\ref{e2}) for different experimental times. The variation of the secret key rate for different times was mainly due to the fluctuation of excess noise induced by atmospheric effects, the change in the probability of channel loss and the variations of FER.

\textit{Conclusion and discussion}.\textemdash
We have demonstrated the first long-distance CVQKD field test over the realistic 7 km inland and 9.6 km maritime atmospheric channels in daytime and nighttime. In particular, we have controlled the total transmission efficiency and excess noise into secure key generation region by developing a series of specialized techniques for free-space CVQKD, including channel-fluctuation-independent high-precision manipulation of continuous-variable quantum states, high-accuracy quantum signal acquisition and processing, and high-efficiency free-space ATP technology. 
Here, we just consider the performance of free-space CVQKD in an asymptotic regime. 
One reason is that the intense atmospheric effects in long-distance free-space channels lead to a limited time window for secure key generation. The other one is that the excess noise suppression techniques in free-space CVQKD system, including the continuous-variable quantum signal manipulation accuracy and stability of photoelectric module and the corresponding software algorithm, are not robust enough, especially for atmospheric channels with medium and strong turbulence. We acknowledge that transitioning from the asymptotic regime to the finite-length regime poses a significant challenge. In the next step, we will improve the free-space system and optimize the grouped subset interval to accumulate more key data to relieve the finite-size effects. 

Based on the estimation of an effective atmospheric thickness of approximately 10 km for satellite-to-ground links \cite{Liaonature2017}, the experimental distance has approximately reached the effective atmospheric range. So it presents a possible alternative way to implement the all-day satellite-based quantum cryptography communication by using Gaussian-modulated coherent states. Moreover, the CVQKD system can be constructed by components well used in classical coherent optical telecommunication systems, which provides a powerful approach to realize secure communication that is naturally compatible with existing ground telecommunication networks. Our results mark an essential step for realizing integrated air-ground quantum access networks and pave the way for cross-domain applications of all-day quantum communications.

We would like to thank H. Zhou, Y. Dou, Y. Li, J. Ren, Q. Zhang, C. Peng and M. Fang for helpful discussions and assistance. This work was supported by the Innovation Program for Quantum Science and Technology (Grant No. 2021ZD0300703), Shanghai Municipal Science and Technology Major Project (2019SHZDZX01), the Key R\&D Program of Guangdong province (Grant No. 2020B0303040002).

\bibliographystyle{apsrev4-2}
\bibliography{apssamp}

\end{document}